\documentclass[preprint2]{aastex}

\usepackage{epsfig}
\usepackage{psfig}
\usepackage{natbib}
\newcommand{\bwp}{PSR~B1957+20}

\usepackage[usenames,dvips]{color}
\usepackage{multirow}

\begin{document}

\title{X-ray studies of the Black Widow Pulsar \bwp}
    
\author{R. H. H. Huang\altaffilmark{1},  A. K. H. Kong\altaffilmark{1}, J. Takata\altaffilmark{2}, 
             C. Y. Hui\altaffilmark{3}, L. C. C. Lin\altaffilmark{4}, K. S. Cheng\altaffilmark{2}}

\altaffiltext{1}
{Institute of Astronomy and Department of Physics, National Tsing Hua University, Hsinchu, Taiwan}
\altaffiltext{2}
{Department of Physics, University of Hong Kong, Hong Kong}
\altaffiltext{3}
{Department of Astronomy and Space Science, Chungnam National University, Daejeon, Korea}
\altaffiltext{4}
{General Education Center of China Medical University, Taichung, Taiwan}

\begin{abstract}
We report on Chandra observations of the black widow pulsar, \bwp. 
Evidence for a binary-phase dependence of the X-ray emission from the pulsar 
is found with a deep observation.
The binary-phase resolved spectral analysis reveals non-thermal X-ray emission of \bwp, 
confirming the results of previous studies. 
This suggests that the X-rays are mostly due to intra-binary shock emission which is strongest 
when the pulsar wind interacts with the ablated material from the companion star. The geometry 
of the peak emission is determined in our study. The marginal softening of the spectrum of the 
non-thermal X-ray tail may indicate that particles injected at the termination shock is dominated 
by synchrotron cooling.
\end{abstract}


\section{Introduction}

The widely accepted scenario for the formation of a millisecond pulsar (MSP) is that an old 
neutron star has been spun up to millisecond periods in a past accretion phase by mass and 
angular momentum transfer from a binary late-type companion \citep{1982alpar,1982radhakrishnan}. 
Once the accretion has stopped, the relativistic magnetized pulsar wind, which is believed to 
carry away the pulsar rotational energy and angular momentum, may be able to ablate and 
eventually evaporate its companion \citep{1989rasio}. Close binary systems 
with MSPs are a subject of special interest since they are thought to be the missing link between 
low-mass X-ray binaries (LMXBs) and isolated MSPs. Discoveries of the eclipsing binary pulsar 
systems, such as \bwp, \citep{1988fruchter}, PSR~J2051--0827 \citep{1996stappers} and PSR J1023+0038
\citep{2009archibald,2010archibald} in the Galactic plane and 47~Tuc~W (PSR~J0024--7204W; 
\citealt{2005bogdanov}) and PSR~J1740--5340 \citep{damico2001b} in globular clusters, gave support 
to this formation scenario. Studying these binary systems provides a wealth of information not only on 
the evolutionary history of isolated MSPs but also on the physical details of the pulsar's high energy 
emission properties. \\[-2ex]

PSR B1957+20 was discovered at Arecibo in 1988 \citep{1988fruchter} . It is in a binary system 
with a 0.025 $\mathrm{M_{\odot}}$ companion in a 9.16-hr orbital period.
The pulsar has a spin period of 1.6 ms, the third shortest among all known MSPs. Its period 
derivative of $\dot{P} = 7.85 \times 10^{-21} ~\mathrm{s~s^{-1}}$ implies a spin-down energy 
of $\dot{E} = 7.48 \pm 3.61 \times 10^{34} ~\mathrm{erg~s^{-1}}$, 
a characteristic spin-down age of $ > 3.4 \times 10^{9}$~yrs, 
and a dipole surface magnetic field of $B_\perp = 1.12 \times 10^{8}$~G \citep{2012guillemot}.
For a radio dispersion measure inferred distance of $2.5\pm1.0$~kpc \citep{2002cordes}, the pulsar 
moves through the sky with a supersonic velocity of 220~km~s$^{-1}$ \citep{1994arzoumanian}. The 
interaction of a relativistic wind flowing away from the pulsar with the interstellar medium (ISM) 
produces an H$\alpha$ bow shock nebula which was the first nebula to be found around a ``recycled" 
pulsar \citep{1988kulkarni}. 
For approximately 10\% of this orbit, the radio emission at a frequency of 430~MHz from the 
pulsar is eclipsed by material ablated by the pulsar wind from the companion \citep{1988fruchter}.
Optical observations by \citet{1988fruchter2} and \citet{1988van} revealed 
that the pulsar wind consisting of electromagnetic radiation and high-energy particles is ablating and 
evaporating its white dwarf companion star. This rarely observed property gave the pulsar the name 
{\em black widow pulsar}. 
Subsequent analyses of the optical light curve for the binary system by \citet{2007reynolds} 
gave a constraint to the system inclination of $65^{\circ}\pm2^{\circ}$ for a pulsar in the mass
range $1.3-1.9~\mathrm{M_{\odot}}$. The effective temperatures of $T=2900\pm110~K$ and 
$T=8300\pm200~K$ for the unilluminated side and the illuminated side of the companion were 
also obtained in their studies. \\[-2ex]

Several X-ray studies of \bwp\ have been carried out in the past two decades 
\citep{1992kulkarni,2003stappers,2007huang}. 
The X-ray emission of \bwp\ is found to be non-thermal dominated and best modeled with a single 
power-law spectrum, which indicates that the X-rays originate from the shock interaction of the pulsar 
wind with the wind of the companion star or from the pulsar magnetosphere \citep{2003stappers,2007huang}. 
Searching for an X-ray pulsation at the radio pulsar's rotation period and a modulation of the X-ray emission 
from the \bwp\ system with orbital phase is another important subject, which may help to discriminate
the origin of the X-ray emission.  
Recently a $4-\sigma$ detection of X-ray coherent pulsation was reported by \citet{2012guillemot}. 
In addition, \citet{2007huang} found a strong correlation of the pulsar's X-ray flux with its orbital period. 
However, due to the short exposure we could not know whether the flux modulation was periodic and 
given the limited photon statistics it was not possible to investigate any spectral variation as 
a function of orbit phase or to determine the exact geometry of the peak emission. Repeated coverage 
of the binary orbit in a longer Chandra observation would provide us a better photon statistic and allow 
us to determine the emission geometry with higher accuracy. \\[-2ex]

In this paper, we report on an archival Chandra observation of the black widow pulsar. 
This investigation offers further insight into the properties of this remarkable binary. 
In \S 2, we summarize the details of the Chandra observation and the data reduction.
A search for the X-ray orbital modulation from this system was described in \S 3. In \S 4 
and \S 5, we present the results of the X-ray spatial and spectral analysis. Finally, we 
discuss the physical implications of the observed results in \S 6.

\section{Observation}

A Chandra observation aimed on \bwp\ was performed on 2008 August 15 (ObsID 9088)
using the back-illuminated chip ACIS-S3 with an uninterrupted 169-ks exposure. The data
was configured in the VFAINT telemetry mode. Data reduction and analysis were processed with 
Chandra Interactive Analysis Observations (CIAO) version 4.3 software and the Chandra 
Calibration Database (CALDB) version 4.4.1. The level 1 data with background cleaning 
were used in our study. Data analysis was restricted to the energy range of $0.3-8.0$\,keV. \\[-2ex]

For the timing and spectral analyses of the black widow pulsar, we extracted the photons from 
a circular region centered at the radio timing position\footnote{from the ATNF Pulsar 
Catalogue}, RA(J2000)=19$^{h}$59$^{m}$36$\fs$77, Dec=20$^{\circ}$48$'$15$\farcs$12, 
with a radius of 2$''$ which encloses 90\% of the total source energy at 1.5\,keV. 


\section{Timing Analysis}

For the timing analysis, we first extracted the photons from the aforementioned circle and 
translated the photon arrival times to the solar system barycenter by using the CIAO tool 
{\em axbary}. The JPL DE200 solar system ephemeris was used for the barycentric 
correction to ensure consistency with the radio ephemeris.
We note that searching for the X-ray pulses at the spin period of \bwp\ was precluded by the 
inappropriate temporal resolution of this Chandra observation with a frame time of 3.2~s. \\[-2ex]

\begin{figure*}
\begin{center}
\includegraphics[width=8.0cm]{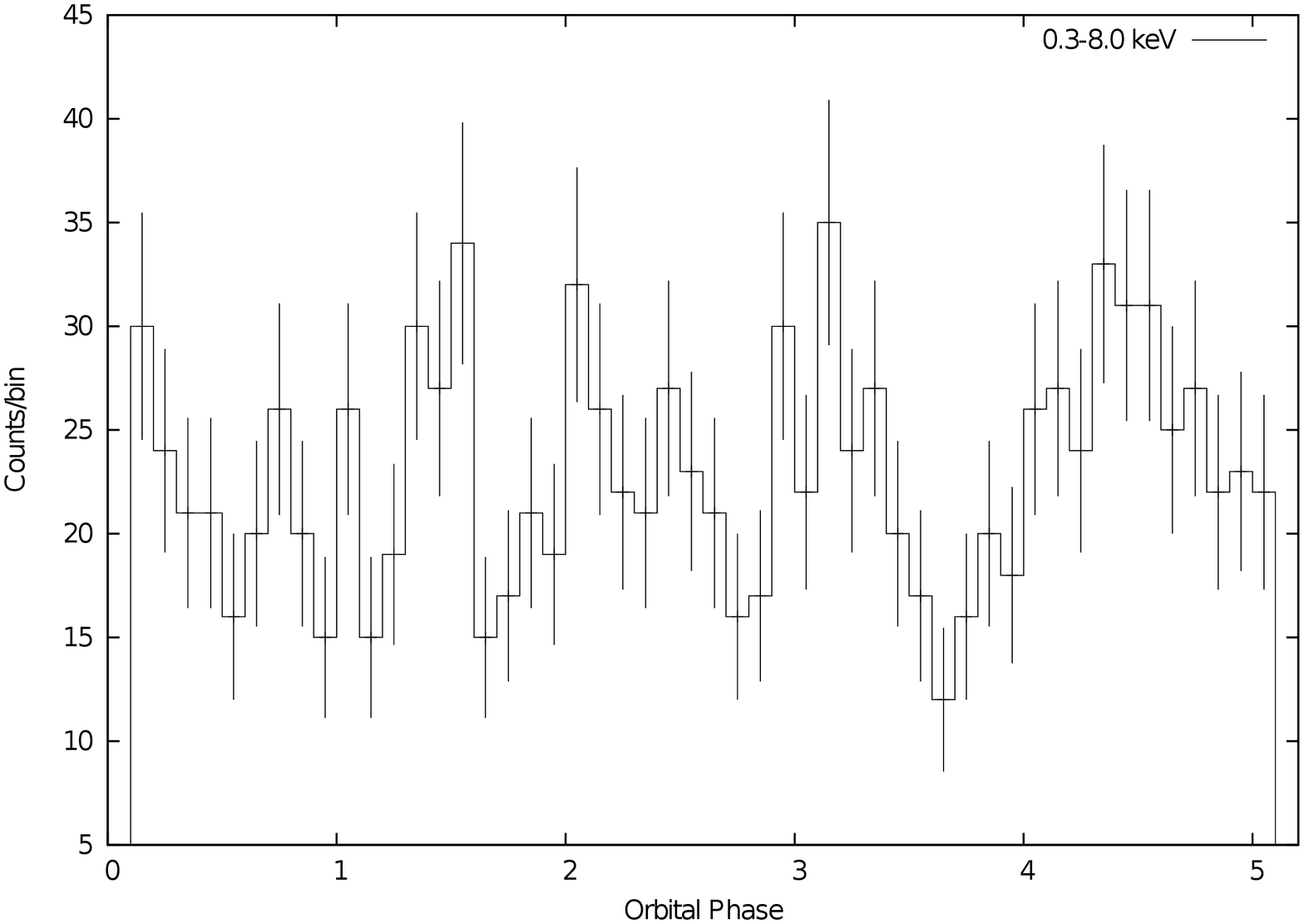}
\includegraphics[width=8.0cm]{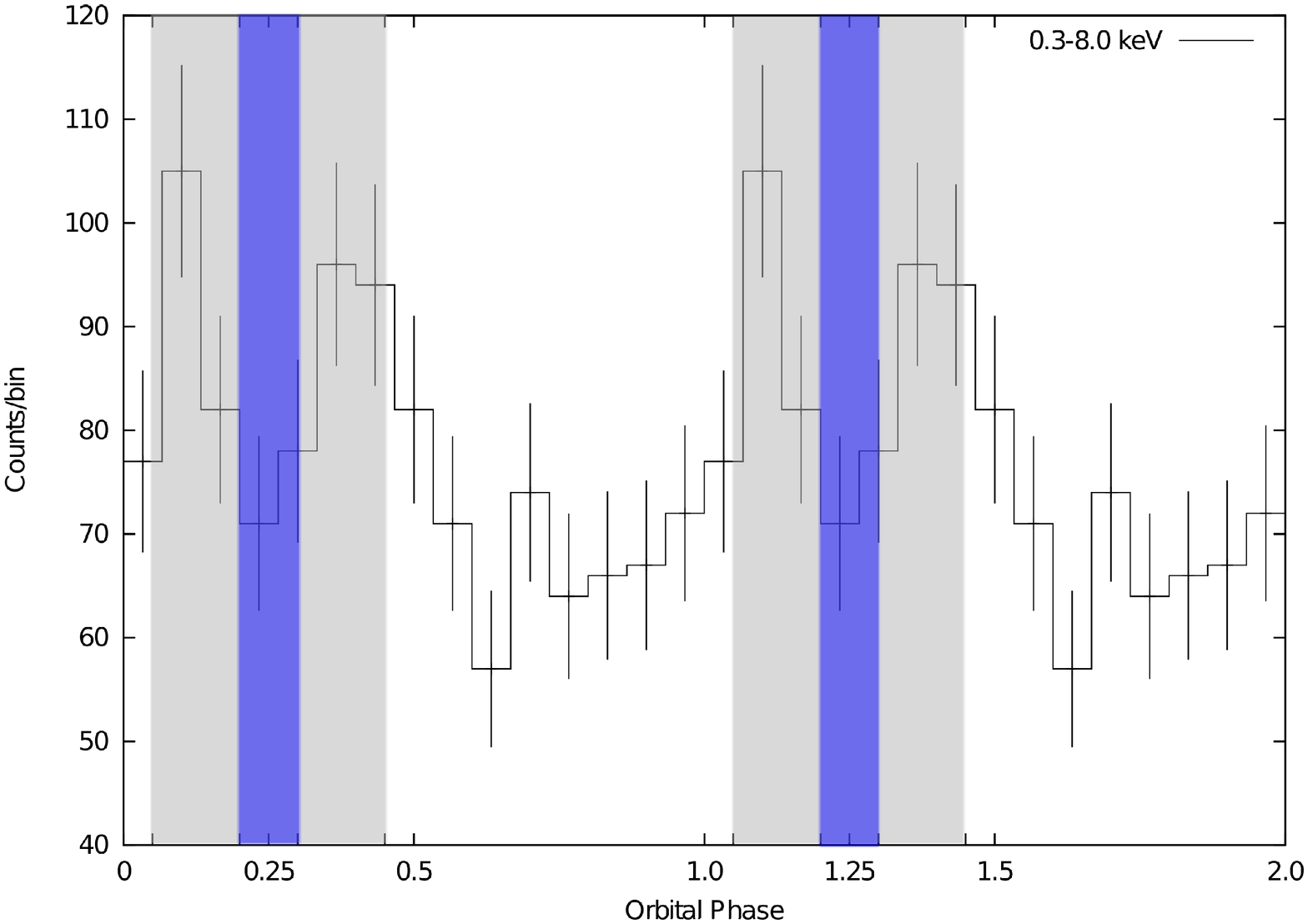}
\caption{ 
{\em Left:} X-ray emission from PSR B1957+20 in the energy range of 0.3--8.0~keV as a 
function of the pulsar’s orbital phase ($\phi$). 
{\em Right:} A folded light curve at the orbital period. Two orbital cycles are shown for clarity. 
The background noise level is found to be at $\sim 0.1$ counts/bin. The phase zero ($\phi$= 0.0) 
corresponds to the ascending node of the pulsar orbit. Error bars indicate the 1$\sigma$ 
uncertainty. The blue shaded  regions between the orbital phases 0.2--0.3 and 1.2--1.3 
mark the radio eclipse of the black widow pulsar. The phase-resolved spectrum covering 
the eclipsing region was extracted from the gray shaded regions (see \S 5.1).}
\label{fig1}
\end{center}
\end{figure*}

As the Chandra observation covers over five consecutive binary orbits, by plotting a light 
curve of the X-ray source counts versus the orbital phase (see Figure~\ref{fig1} left panel) 
we can confirm that the X-ray flux is not steady with time. 
We also applied a Kolmogorov-Smirnov (KS) test to the unbinned light curve data in 
order to have a bin-independent statistical evaluation of the X-ray emission variability.
Calculating the corresponding KS probabilities between our data set and the cumulative 
distribution function generated by assuming a constant X-ray flux indicated a marginal 
deviation between these two distributions. The significance for an intra-orbital flux 
modulation from this test is only at the level of $\sim$81\%. A more significant result, 
i.e. $\sim$96\%, is obtained if we restrict the analysis to the soft energy band below 
2~keV. \\[-2ex]

In order to search for a modulation of the X-ray flux as a function of orbital phase, 
we first selected X-ray data covering 5 complete and consecutive orbits and then 
used the radio timing ephemeris of \bwp\ from a pulsar catalog provided by Lucas 
Guillemot\footnote{ftp://www.cenbg.in2p3.fr/astropart/lucas/report/1959+2048.html}
to fold a light curve at the orbital period (see Figure~\ref{fig1} right panel). Using a 
$\chi^{2}$-test, the significance for a flux modulation over the observed orbit was 
found to be $\sim$ 99\%.

\section{Spatial Analysis}

\begin{figure*}
\centerline{\psfig{figure=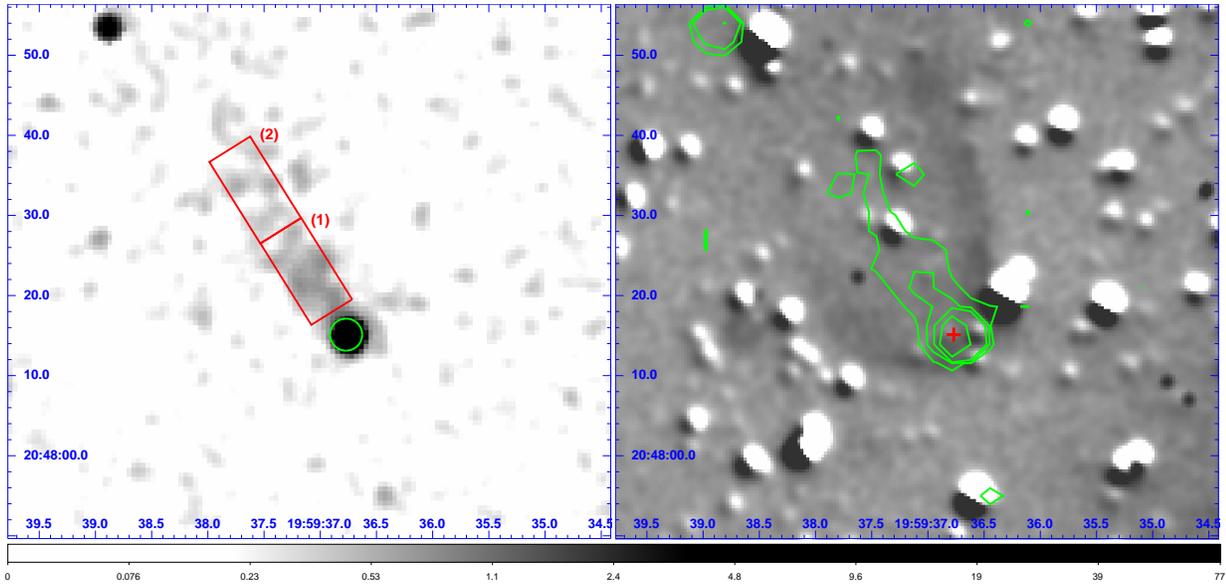,width=17cm,clip=}}
\caption{{\em Left panel:} Chandra ACIS-S3 image in the energy band 0.3--8~keV of the 
black widow pulsar system smoothed with an adaptive Gaussian filter. 
The green circle with the $2.0''$ radius indicates the source region we used in this study
while two segments of the X-ray tail are chosen from the red rectangular regions.
{\em Right panel:} The H$\alpha$ image taken from the Anglo Australian Telescope is overlaid 
with the X-ray contours. The green contour levels are shown at 0.4, 0.8, 3.0, 15.4, and 84.8\% 
of the peak x-ray surface brightness. The red cross indicates the radio timing position of \bwp.
The optical residuals correspond to incompletely subtracted stars. 
\label{fig2}}
\end{figure*}

Figure~\ref{fig2} (left panel) shows the Chandra ACIS-S3 image in the energy band 0.3--8~keV of the field 
around \bwp. This image was created by using an adaptive smoothing algorithm with a Gaussian 
kernel of $\sigma < 3$ pixels in order to probe the detailed structure of faint diffuse emission. Both the pulsar 
and an extended X-ray feature (hereafter the ``tail''), protruding from the pulsar position, can be clearly seen 
in this image. The length of the tail with its orientation to the northeast is about 25 arcsec. 
The right panel in Figure~\ref{fig2} presents the H$\alpha$ image overlaid with the X-ray contours. 
This H$\alpha$ image is obtained from Taurus Tunable Filter service mode observations on the 
Anglo Australian Telescope in 2000. An H$\alpha$ bow-shock emission aligned with the proper 
motion direction of the \bwp\ system fades to the background in $\sim 40''$. Comparing with 
the previous result reported by \citet{2003stappers}, we found that the X-ray tail is more extended 
 due to the longer exposure time of this Chandra observation.

\section{Spectral Analysis}

\subsection{\bwp}

We used the CIAO tool {\em dmextract} to extract spectra of the source and two nearby source-free 
background regions. Response files were constructed by using the CIAO tool {\em mkacisrmf} and 
{\em mkarf}. The extracted spectra were binned with at least 30 source counts per bin. Background-subtracted 
spectral modeling was performed with XSPEC (version: 12.7.0) using data in the energy band 0.3-–8.0~keV. \\[-2ex]

Assuming that the X-ray emission originates from the intra-binary shock or the pulsar's magnetosphere
\citep{2003stappers,2007huang,2012guillemot}, we expect the radiation to be synchrotron. 
To test this hypothesis, we fitted the spectrum with an absorbed power-law model (PL). Unexpectedly, 
a single PL model cannot provide any statistically acceptable description of the observed spectrum 
(i.e. $\chi^{2}_{\nu} > 2$). We then tested whether a single blackbody (BB) model, a double PL model, 
or a composite spectral model consisting of a PL and a BB can provide an appropriate modeling of the data. 
Neither the one-component BB model nor the double-component PL model yields any physically acceptable fits.
Instead, the composite model gives a better description of the X-ray spectrum. The inferred temperature 
of 0.18~keV and 0.30~keV is much too high for the cooling surface of the neutron star, but may be 
explained by the heated polar cap region (c.f. \citealt{2007zavlin,2012takata}). 
Considering that a portion of the X-ray emission could originate from a thermal plasma within or around 
the binary system, possibly from the active corona of the secondary star or the plasma responsible for the 
radio eclipses \citep{2011bogdanov}, a MEKAL, a thermal bremsstrahlung (TB), a MEKAL+PL, and 
a PL+TB model were tested in our study. We found those models cannot provide a better or an acceptable 
description of the X-ray spectrum of the pulsar system. Besides, we also examined the X-ray spectrum with more 
complicated spectral models, e.g. PL+PL+BB or PL+BB+BB. Although an acceptable fit can be yielded by
a three-component model, i.e. PL+BB+BB, no physical interpretation can be appropriately applied to this binary 
system. Therefore, we suspect a dependence of the X-ray spectrum of \bwp\ on its orbital phase due to 
the variability observed in its X-ray flux level which seems to correlate with its orbital period. \\[-2ex]

To investigate whether the X-ray spectral behavior of \bwp\ varies across the orbit, we analysed the X-ray 
spectra within the orbital phase of $\phi = 0.05 - 0.45$ which covers the ingress, eclipsing, and egress
region and outside the aforementioned region ($\phi = 0.45 - 1.05$) separately. The source and background 
spectra were extracted from the same circles as mentioned before and dynamically binned in accordance 
with the photon statistic in each dataset. The net count rates for the spectral analysis inside and outside 
the eclipsing region are $7.49 \pm 0.33$ and $5.88 \pm 0.25$ counts per kilosecond, respectively. 
We found that the binary-phase resolved spectral analysis reveals a non-thermal emission nature of the 
detected X-rays and each of the observed spectra can be well described by a single PL model with different 
photon indices, which indicates that its spectral behavior is orbital dependent. \\[-2ex]

The spectral parameters inferred from the PL fits as well as the best-fit hydrogen column density 
$\mathrm{N_{H}}$ are found to be consistent with those reported by \citet{2003stappers} and 
\citet{2007huang}. For comparison, a fixed $\mathrm{N_{H}}$ value of $1.8\times10^{21}$~cm$^{-2}$ 
obtained by \citet{2003stappers} is also applied in our spectral analysis. 
The results of the spectral fits are summarized in Table~1. 

\subsection{The X-ray Tail}

\citet{2003stappers} uncovered an X-ray tail with a position angle opposite to the proper motion direction 
of \bwp\ with a Chandra observation taken in 2001. However, with low photon statistics, a detailed spectral 
analysis was restricted. The much longer exposure time of the latest archival Chandra observations 
allows us to study the X-ray tail in greater detail.
For comparison, we first selected a box of 16$''\times$6$''$ with an orientation along the proper 
motion direction as the region of the X-ray tail of \bwp. The background-subtracted count of $108\pm13$ 
was collected for the spectral analysis. With higher photon statistics, we found an absorbed single PL model 
fits the X-ray spectrum of the tail well, which implies that the X-ray emission originates from the pulsar’s 
interaction with the ambient medium. Its non-thermal X-ray spectrum points to synchrotron emission from 
energetic particles from the pulsar wind. \\[-2ex]

A softening of the spectrum of the X-ray tail as a function of the distance from the pulsar is expected 
if synchrotron cooling of the particles injected at the termination shock is dominated. For the purpose 
of investigating the possible spectral variation, we performed a spatially-resolved spectral analysis using 
two separate extraction regions along the tail. We refer the segment close to the pulsar as region 1 
and the further one as region 2 in the following. The background-subtracted counts are $68\pm14$ for 
region 1 and $29\pm12$ for region 2, respectively. In order to better constrain the spectral properties, 
we followed the method adopted by \citet{2010johnson} and \citet{2012hui} to jointly fit individual 
power-law models for the X-ray spectra of these two regions assuming the column density, 
$\mathrm{N_{H}}$, does not change significantly along the proper motion direction. Due to low photon 
statistics in each segment, we then fixed the $\mathrm{N_{H}}$ value at $1.8\times10^{21}$~cm$^{-2}$ 
\citep{2003stappers} in the joint fit. The best-fit spectral parameters are shown in Table~1. An indication 
for such a spectral variation was found in this study. \\[-2ex]

\begin{center}
\begin{deluxetable}{lccccc}
\tablewidth{0pc}
\tablecaption{X-ray spectral parameters of the \bwp\ binary system}
\startdata
\hline\hline
Orbital Phase & Model & N$_\mathrm{H}$ & $\Gamma$/kT & $\mathrm{F_{x}}$\tablenotemark{a} & $\chi^{2}_\nu$/d.o.f  \\
                     & & ($10^{21}$~cm$^{-2}$) &  /(keV)& ($10^{-14}$~ergs~cm$^{-2}$~s$^{-1}$) &            \\ \hline\hline \\[-2ex]
\multicolumn{6}{c}{Pulsar}\\ \hline \\[-2ex]
0.0-1.0       & PL & $0.83^{+0.33}_{-0.32}$ & $1.96\pm0.12$             & $6.05^{+0.26}_{-0.28}$ & 1.86/18  \\
                   & PL & 1.80\tablenotemark{b}   & 2.26\tablenotemark{c}   & $7.22^{+0.13}_{-0.14}$ & 2.17/19   \\ 
                   & PL+BB & $<0.61$   & $\Gamma$=$1.02^{+0.37}_{-0.31}$, kT=$0.30^{+0.02}_{-0.05}$   & $5.39^{+0.06}_{-1.14}$ & 1.48/16   \\ 
                   & PL+BB & 1.80\tablenotemark{b}   & $\Gamma$=$1.66^{+0.22}_{-0.25}$, kT=$0.18^{+0.02}_{-0.03}$  & $7.18^{+0.12}_{-0.28}$ & 1.75/17   \\ \\[-2ex] 
\hline \\[-2ex]
0.05-0.45   & PL & $0.78^{+0.56}_{-0.53}$ & $1.80^{+0.19}_{-0.17}$ & $8.05^{+0.53}_{-0.70}$ & 0.75/13   \\
                   & PL & 1.80\tablenotemark{b}   & $2.09\pm0.10$             & $9.23^{+0.43}_{-0.39}$ & 0.92/14    \\ [0.5ex]
\hline \\[-2ex]
0.45-0.05   & PL & $0.89^{+0.54}_{-0.50}$ & $2.17^{+0.21}_{-0.20}$ & $6.12^{+0.36}_{-0.38}$ & 0.99/15   \\
                   & PL & 1.80\tablenotemark{b}   & $2.49^{+0.11}_{-0.10}$ & $7.61^{+0.24}_{-0.19}$ & 1.10/16    \\  [0.5ex]
\hline\hline
\\
\\
\hline\hline 
Segment      & Model & N$_\mathrm{H}$ & $\Gamma$ & $\mathrm{F_{x}}$\tablenotemark{a} & $\chi^{2}_\nu$/d.o.f  \\
                   & & ($10^{21}$~cm$^{-2}$) &       & ($10^{-14}$~ergs~cm$^{-2}$~s$^{-1}$) &            \\ \hline\hline \\[-2ex]
\multicolumn{6}{c}{Tail}\\ \hline \\[-2ex]
Whole tail    & PL & $2.77^{+3.21}_{-2.42}$ & $2.62^{+1.23}_{-0.84}$  & $0.89^{+0.10}_{-0.19}$ & 0.77/4     \\
                   & PL & 1.80\tablenotemark{b}   & $2.30^{+0.39}_{-0.36}$  & $0.71^{+0.10}_{-0.07}$ & 0.64/5    \\ 
\hline \\[-2ex]              
Region 1     & PL  & \multirow{2}{*}{1.80\tablenotemark{b}}        & $1.57^{+0.31}_{-0.26}$  & $0.46^{+0.13}_{-0.10}$ & \multirow{2}{*}{1.28/18}    \\
Region 2     & PL  &    & $2.14^{+0.46}_{-0.37}$  & $0.33^{+0.16}_{-0.05}$   &     \\  [0.5ex]                
\hline
\enddata
\tablenotetext{a}{Unabsorbed X-ray flux in the energy range of 0.3--8.0 keV.}
\tablenotetext{b}{The hydrogen column density N$_\mathrm{H}$ is fixed at $1.8\times10^{21}$~cm$^{-2}$.}
\tablenotetext{c}{No error was calculated since the reduced $\chi^{2}$ is larger than the maximum value of 2.}
\label{spec_par}
\end{deluxetable}
\end{center}


\section{Summary \& Conclusion}

We have searched for the orbital modulation of the X-ray emission from \bwp. Analysing 
this data set with a $\chi^{2}$-test and a Kolmogorov-Smirnov test revealed a marginal  
intra-orbital flux modulation, which suggests that the non-thermal X-rays from \bwp\ are 
mostly due to intra-shock emission at the interface between the pulsar wind and the 
ablated material from the companion star. The pulsar wind electrons and positrons are 
accelerated and randomized by the shock and emit the X-rays via the synchrotron process. 
Such shock emission has also been suggested to explain the variable X-ray flux as 
a function of orbital phase in 47~Tuc~W \citep{2005bogdanov} and PSR~J1740--5340 
in NGC~6397 \citep{2010huang} and PSR~J1023+0038 \citep{2009archibald,2010archibald,
2010tam,2011bogdanov}.

In Figure~\ref{fig1}, the observed flux peaks just before and after the pulsar eclipse can be 
interpreted as the Doppler effect caused by the bulk flow in the down stream region. 
If the post-shocked wind flows toward (or away from) the Earth, the Doppler effect increases 
(or decreases) the observed flux  from the flux for an isotropic case. The shock geometry is 
controlled by the ratio of the momentum fluxes of the pulsar wind to the stellar wind 
(c.f. \citealt{1996canto,2004antokhin}). For PSR~B1957+20, the observed orbital 
period derivative $\dot{P}_b\sim 10^{-11}$ \citep{1990fruchter} suggests the mass loss rate  
$\dot{M}_*\sim M_{*}\dot{P}_{b}P_b^{-1}\sim 10^{-10}~\mathrm{M_{\odot}~yr^{-1}}$, 
where $M_{*}=0.02~\mathrm{M_{\odot}}$ and $P_{b}=33001$~s is the orbital period. 
With the mass-loss rate of $\dot{M}\sim 10^{-10}~\mathrm{M_{\odot}~yr^{-1}}$,
the ratio of the momentum fluxes of the pulsar wind and the stellar wind is in the order of 
$\eta\equiv L_{sd}/\dot{M}_*v_{es} c\sim 10 (L_{sd}/10^{35}~\mathrm{erg~s^{-1}})
(\dot{M}_*/10^{-10}~\mathrm{M_{\odot}~yr^{-1}})(v_{es}/300~\mathrm{km~s^{-1}})$, 
indicating  the companion star is confined by the pulsar wind and the shock. 
With momentum ratio $\eta\sim 10$, the opening angle of the cone-like shock is 
$\sim 50-60^{\circ}$ (c.f. \citealt{1993eichler,1996canto}), which corresponds to 
$\sim 0.15$ orbital phase. Because the emission is concentrated in the forward direction 
of the flow, therefore, we expect that double peaks due to the  Doppler effect appear at 
$\sim 0.15$ phase before and after the phase of radio eclipse, which is consistent with 
the observation. \\[-2ex]

As we can see in Figure~\ref{fig1}, the observed
 ratio of  maximum to minimum fluxes is $2\sim 3$.  
For the emissions from the pulsar wind, 
the outgoing flux is modified by the Doppler effect as 
$F_{\nu}(E)={\mathcal D}^3F'_{\nu'}(E')$ and $E={\mathcal D}E'$, 
with primed quantities refereeing to the comoving frame. 
 Here the relativistic boosting factor is given by 
\begin{equation}
{\mathcal D}=\frac{1}{\Gamma_f (1-\beta_f\cos\theta_f)}, 
\label{dop}
\end{equation}
where $\Gamma_f$ and $\beta_f$ is 
the Lorentz factor and  velocity in units of the speed of light of the flow, 
respectively,  and $\theta_f$ is the angle between the Earth viewing angle and the direction of the flow. The 
ratio of maximum to minimum fluxes  is estimated by 
\begin{equation}
\frac{F_{\nu, max}}{F_{\nu, min}}
\sim \left(\frac{1-\beta_f\cos\theta_{f,min}}{1-\beta_f\cos\theta_{f,max}}\right)
^{2+\alpha},
\label{ratio}
\end{equation}
where $\alpha$ is the photon index, $\theta_{f,max}$ and
 $\theta_{f,min}\sim \pi-\theta_{f,max}$  are 
typical angles at the orbital phases where the observed fluxes are maximum
 and minimum, respectively. For example, the observed 
ratio $F_{max}/F_{min}\sim 2-3$ and the photon index $\alpha\sim 2$
 implies the flow velocity $\beta_f\sim 0.12-0.19$ for 
$\theta_{f,max}=45^{\circ}$ and $\beta_s\sim 0.2-0.32$ for 
$\theta_{f,max}=65^{\circ}$. Note that \citet{2012guillemot} estimates 
the fraction of the pulsed emissions as about $30$~\% of the total emissions. 
In such a case, the pulsed emissions may considerably contribute to
 the observed emissions at off-peak orbital phase, and 
the intrinsic ratio of the maximum to minimum fluxes of the 
X-ray emissions from the inter-binary shock  is larger than 
$F_{max}/F_{min}\sim 2-3$. This results in increase in the  
flow velocity estimated from equation~(\ref{ratio}). \\[-2ex]

Although the pulsed X-ray emission from PSR~B1957+20 has been reported 
by \citet{2012guillemot}, the origin of the pulsed emissions is not known. 
As Table~1 shows, we find that the spectrum average over the whole orbit can be fitted
better by a power law plus blackbody model. The effective temperature and radius of 
the black body component are $T_{eff} \sim 2\times 10^6$~K and $R_{eff} \sim 0.1$~km, 
respectively, which are explained by  the core component of the heated polar cap region 
(c.f. \citealt{2007zavlin,2012takata}). The observed flux of the blackbody component  is 
$F_{BB}\sim 2\times 10^{-14}~\mathrm{erg~cm^{-2}~s^{-1}}$, which is several tens of 
percent of the total emissions. This flux level of the pulsed emission is consistent with the 
result obtained by Guillemot et al. (2012), who have estimated the fraction of the pulsed 
emissions as about $30$~\% of the total emissions. \\[-2ex]

X-ray tails around pulsars have been interpreted as bow-shocks generated by the supersonic 
motion of pulsars through space, with the wind trailing behind as its particles are swept back 
by the pulsar's interaction with the interstellar gas it encounters. Studying the diffuse X-ray 
emission may help to better understand their ambient environment and the interaction
between the pulsar and the ISM. Until now only two MSPs, \bwp\ \citep{2003stappers} and 
PSR J2124--3358 \citep{2006hui}, are found to be associated with extended X-ray emission. 
It is worth to revisit \bwp\ with a deep Chandra observation. 
A $\sim 25''$ X-ray tail extending from \bwp\ with a position angle opposite to the pulsar's 
proper motion direction is clearly resolved in the Chandra ACIS image. The non-thermal 
nature and a marginal softening of the spectrum of the tail in X-rays as a function of the 
distance from the pulsar support the scenario that particles injected at the termination shock 
is dominated by synchrotron cooling. \\[-2ex]

\citet{2006cheng} suggests that the observed length ($l$) of the X-ray tail can be interpreted 
as the distance traversed by the pulsar within the electron synchrotron cooling timescale ($t_{c}$), 
i.e., $l \sim v_{p}t_{c}$, where $v_{p}$ is the proper-motion velocity of the pulsar. The cooling 
time in the X-ray band is $\sim 10^{8} B_\mathrm{mG}^{-3/2}(h\nu_\mathrm{X}/\mathrm{keV)}^{-1/2}$~s, 
where $B_\mathrm{mG}$ is the inferred magnetic field strength in the emitting region \citep{2006cheng}. 
Adopting the inferred tail length of $\sim 9.7\times10^{17}$~cm at the distance of 2.5~kpc 
\citep{2002cordes} and a proper motion velocity of 220~km~s$^{-1}$ \citep{1994arzoumanian}, 
the synchrotron cooling timescale is estimated to be $\sim 4.9\times 10^{4}$~yrs. This yields 
a magnetic field of $B \sim 17.7 ~\mu$G in the shock region. Considering a magnetic field strength 
of $\sim 2-6 ~\mu$G in the ISM (cf. \citealt{2003beck}, and references therein), we found that the 
magnetic field in the termination shock might be compressed by a factor of $\sim 3$, which is consistent 
with the estimated value reported by \citet{1984kennel}.

\acknowledgments
This project is supported by the National Science Council of the Republic of China
(Taiwan) through grant NSC100-2628-M-007-002-MY3 and NSC100-2923-M-007-001-MY3.
AKHK gratefully acknowledges support from a Kenda Foundation Golden Jade Fellowship.
CYH is supported by the National Research Foundation of Korea through grant 2011-0023383.
KSC is supported by the GRF Grants of the Government of the Hong Kong SAR under HKU 7009/11P.
LCCL is supported by the National Science Council of the Republic of China (Taiwan) through 
grant NSC101-2112-M-039-001-MY3.

\bibliography{bibmain}

\clearpage

\end{document}